# Deep Subwavelength Electromagnetic Transparency through Dual Metallic Gratings with Ultranarrow Slits

Chunyin Qiu[1]*, Sucheng Li[2], Ruirui Chen[2], Bo Hou[2]*, Feng Li[3], and Zhengyou Liu[1]

[1]Key Laboratory of Artificial Micro- and Nano-structures of Ministry of Education and School of Physics and Technology, Wuhan University, Wuhan 430072, China

[2]School of Physical Science and Technology, Soochow University, Suzhou 215006, China

[3]Changchun Institute of Optics, Fine Mechanics and Physics, Chinese Academy of Sciences, Changchun 130033, China

**Abstract:**

In this Letter, we study the transmission response of microwaves through two identical metallic plates machined with ultranarrow slit arrays. The measured and calculated transmission spectra consistently display a striking transmission peak at wavelength much larger than any characteristic length of the structure (e.g., about twenty-fold of the lattice period), which can not be directly explained by the existing mechanisms. Both the LC-circuit-based microscopic picture and the effective-medium-based macroscopic model are established to capture the essential physics behind such unexpected resonance at the deep subwavelength scale. Prospective applications of this novel transmission property can be anticipated, considering the merits of compact and excellent immunity to structural imperfections.

In recent years, the extraordinary optical transmission (EOT) [1-7] through an opaque metallic plate perforated with subwavelength apertures has drawn tremendous attention because of many fascinating applications. The studies involve various grating geometries and span a broad range of frequency regimes (from microwaves to optic waves) [8]. Now it is widely accepted that the exotic transmission enhancement can be attributed to the resonances induced by the coupling of the external electromagnetic (EM) waves with apertures, either individually or collectively: the former stems from the Fabry-Perot (FP) resonance of the fundamental waveguide mode inside the slit, in which the resonant wavelength is determined by the sample thickness [4-7]; the latter is induced by lattice resonance and always accompanied with exciting bound states (either the intrinsic surface plasmons or the structure-induced spoof surface waves) on the metal surface [1-3,7], where the resonant wavelength is comparable with the structure period unless the plasma frequency (in ultraviolet regime for ordinary metals) is approached. The continuum crossover connecting these two mechanisms has also been pointed out for one-dimensional (1D) periodical slit arrays [7].

The aim of this Letter is to propose a novel physics mechanism responsible for the EOT effect. The system under consideration consists of two identical metallic plates perforated with ultranarrow slit arrays. The transmission spectra of such dual metallic gratings (DMGs) show a remarkable peak at wavelength much larger than *any characteristic length* in the structure, which is absent in single metallic grating (SMG) systems. The resonant frequency can be captured well by an equivalent LC-Circuit. Considering the characteristics of deep subwavelength in such resonances, we have further developed an analytic model, in which each SMG is macroscopically treated as a homogeneous metamaterial slab with exotic dielectric properties, to reveal the essential physics behind the resonances. Throughout this Letter, the finite-element-based software package (Comsol Mutiphysics 4.2) is adopted for full-wave calculations, and the dedicated microwave experiment is performed to measure the transmission spectra. The experimental results agree excellently with the numerical ones.

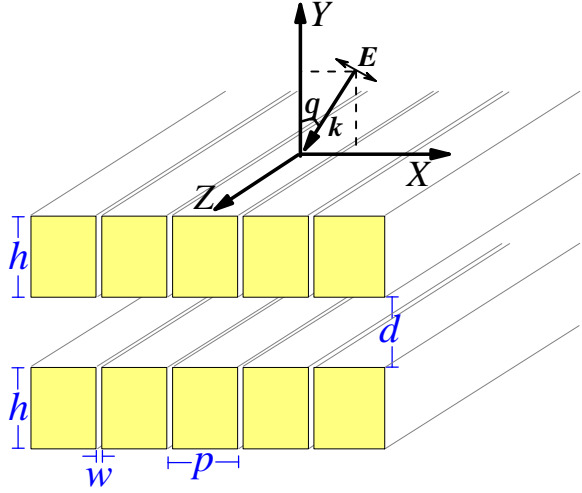

FIG. 1 (color online). Schematic view of the sample illuminated by a microwave of TM-polarization with magnetic field along the *z*-axis and electric field in *x-y* plane. Besides the adjustable geometry parameter $d$, the other lengths are $p = 5.0$mm, $h = 5.8$mm, and $w = 0.4$mm, respectively.

As schematically illustrated in Fig. 1, each SMG is an aluminum plate (of thickness $h = 5.8$mm) machined with a 1D periodical (period $p = 5.0$mm) array of cut-through slits (of width $w = 0.4$mm). The narrow slits are prepared by using the commercial wire-cut electrical discharge machining. The DMG forms when the two identical SMGs are perfectly aligned and separated by air spacer with precisely controllable distance $d$. In the experiment, the SMG or DMG sample is placed in between a couple of collinearly aligned rectangular microwave horns, one acting as the signal generator and the other as a receiver. The structure is illuminated by a microwave of transverse-magnetic (TM) polarization at incident angle $q$, where the magnetic field acts along the *z*-axis, and the electric field lies in the *x-y* plane. The transmission spectrum is determined by the ratio between the transmitted signals with and without the sample in presence.

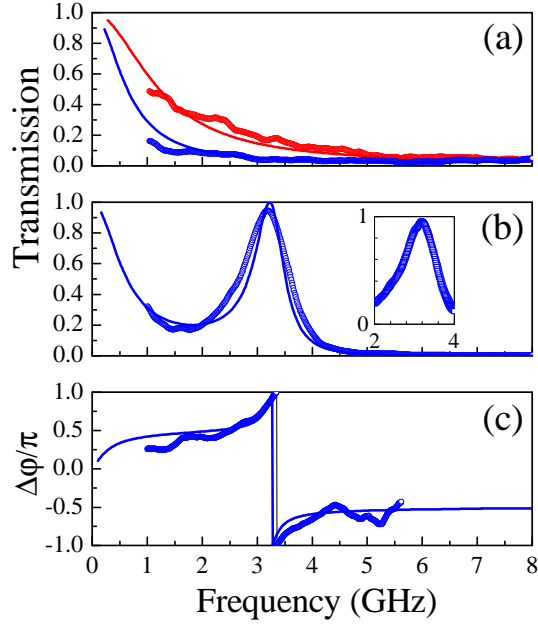

FIG. 2 (color online). The measured (circle) and calculated (line) transmission spectra at normal incidence for the cases of (a) SMG (red) and DMG (blue) with $d=0$, and (b) DMG with $d=5.0$mm, where the inset in (b) exhibits the coincidence of the measured spectra for the cases without (circle) and with lateral shift of half period (square). (c) The measured (circle) and calculated (line) phase accumulation $\Delta\varphi$ across the DMG with $d=5.0$mm.

The transmission of microwaves through a simple SMG structure with narrow slits has been well explored. It is expected to be considerably low over a wide range of the low frequency, where both the localized FP resonance and the collective lattice resonance can not occur. The conclusion also holds for the DMG with zero separation, essentially a SMG of thickness $2h$. These are indeed manifested in Fig. 2(a) consistently by the experimental and numerical transmission spectra at normal incidence. However, when the separation $d$ of the DMG is tuned to 5.0mm, as shown in Fig. 2(b), a striking peak appears around 3.2GHz in the experimental transmission spectrum, in excellent agreement with the numerical prediction. The corresponding resonant wavelength, $l_R=93.2$mm, is much larger than any characteristic length in the DMG structure ($l_R/w$; 230, $l_R/h$; 16, and $l_R / p$; 19), precluding the possibilities of the collective lattice resonance and the individual FP resonance (in the usual sense). We have also measured the transmission spectra for the DMG with relative lateral shifts (along $x$ direction) between the two SMGs. An example associated with a shift of half period is presented in the inset of Fig. 2(b). It is observed that the peak position is almost unaffected, compared to the situation of perfect alignment. This stability can be understood later by an effective-medium-based macroscopic picture with fine structural features neglected. In Fig. 2(c) we present frequency dependence of the phase accumulation $\Delta\varphi$ across the DMG sample. It is surprising that although the whole sample is considerably thin (with respect to $l_R$), the phase difference grows rapidly to $\pi$ near the resonant frequency.

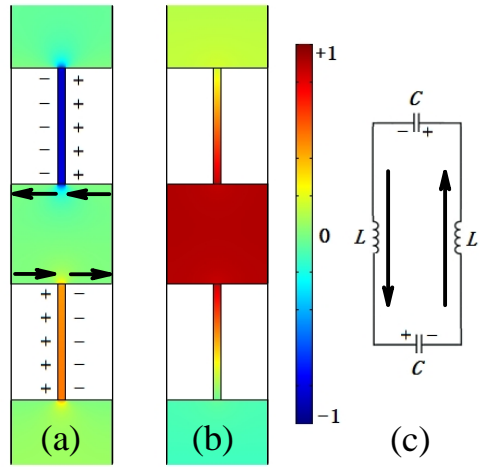

FIG. 3 (color online). Instantaneous $E_x$ (a) and $H_z$ (b) field distributions at the resonant frequency, normalized by the corresponding maximum values and sharing the same color bar. (c) The equivalent LC-circuit that accounts for the surface charge (sign) and current (arrow) depicted in (a).

In order to gain physics insight of the aforementioned EM transparency emerging at deep subwavelength, in Figs. 3(a) and 3(b) we present the instantaneous field distributions for the electric component $E_x$ and the magnetic component $H_z$, obtained by the full-wave calculation at the resonant frequency. It is observed that $E_x$ is strongly squeezed inside the narrow slits (associated with enhancement over tenfold), whereas $H_z$ is mostly concentrated in the air space sandwiched between the two conductive gratings. This alternating EM field induces the oscillation of the free charges and currents on the highly conductive surfaces, as displayed schematically in Fig. 3(a). The oscillatory surface charge and current configurations can be accounted for by an equivalent LC-circuit [9,10] manifested in Fig. 3(c). For simplicity, here the dissipative resistor elements (which are negligibly small in the microwave regime) are omitted. Since the dominant electric (or magnetic) field is distributed almost uniformly in the slits (or the sandwiched air cavity), each capacitance (or inductance) can be estimated by $C = e_0 h \Lambda / w$ (or $L = 0.5 m_0 dp / \Lambda$), where $e_0$ and $m_0$ are respectively the electric permittivity and magnetic permeability in vacuum, and $\Lambda$ is the length along the $z$-axis (assumed to be infinity). The complex impedance of the whole LC-circuit is $Z_c = 2[iwL + (iwC)^{-1}]$, with $w$ being the angular frequency. By requiring $Z_c = 0$ one can obtain a resonant frequency $F_R = \frac{c_0}{2p}\sqrt{\frac{2w}{phd}}$, at which the EM wave can be efficiently harvested and subsequently squeezed through the double slit arrays. Here $c_0 = 1/\sqrt{m_0 e_0}$ is the light speed in vacuum. When the slit is sufficiently narrow, the corresponding resonant wavelength (in vacuum) $l_R = p\sqrt{2phd/w}$ can be much larger than the other geometric lengths, i.e., the structure period $p$, grating thickness $h$, and separation $d$. For the current DMG system with $d = 5.0mm$, the resonant wavelength $l_R$ predicted by the simple LC-circuit model is $84.6mm$, deviating by less than 10% from the above full-wave calculation, i.e., $93.2mm$.

In fact, the above EOT effect can also be understood in a macroscopic way. At wavelength much larger than the microstructures of the sample, each SMG (made of perfect conductor) can be viewed as a homogenous metamaterial plate of thickness $h$, characterized by extremely anisotropic electric permittivity and magnetic permeability tensors [11]. Both tensors are controlled only by the filling ratio of the slit $\eta = w/p$. The effective-medium-theory (EMT) model is particularly valid when the slit is long and narrow (i.e., $w \ll h$), where the evanescent nonzero order of waveguide mode inside the slit can be safely omitted. For clarity, we first focus on the normal incidence that involves only the *x*-component of permittivity $\varepsilon_x = \varepsilon_0/\eta$ and the *z*-component of permeability $\mu_z = \eta\mu_0$. It is of interest that such exotic material parameters keep the propagation constant $k_0$ invariant, but give a tiny impedance ratio $Z_r = \eta$ with respect to free space, which are essentially inaccessible in natural materials. For the artificial bilayer structure it is easy to analytically derive the reflection and transmission coefficients by using the conventional transfer-matrix method. The reflection coefficient exhibits two distinct zero reflection conditions. The first one is determined by $k_0 h = m\pi$ (with $m$ being positive integer) and corresponds to the usual FP resonance in each artificial dielectric plate. It is only dependent on the sample thickness and has been extensively studied in previous SMG systems [4-7]. The second one is determined by the formula

$$2Z_r / (1+Z_r^2) = \tan(k_0 h)\tan(k_0 d), \tag{1}$$

which is unique in the bilayer system. In view of the characteristic of the trigonometric function, Eq. (1) has a series of solutions. Here we focus on the fundamental one (associated with the lowest resonant frequency). For narrow slit case (with $Z_r = \eta \ll 1$), Eq. (1) can be well approximated by $2Z_r \cong \tan(k_0 h)\tan(k_0 d)$. At the low frequency (i.e., $k_0 d \ll 1$ and $k_0 h \ll 1$), it is further simplified into $2\eta \cong k_0^2 hd$ considering that $\tan(k_0 d) \cong k_0 d$ and $\tan(k_0 h) \cong k_0 h$. This gives rise to the fundamental resonant frequency [12]

$$F_R \cong \frac{c_0}{2\pi}\sqrt{\frac{2\eta}{hd}}, \tag{2}$$

which is exactly the one deduced from the LC-circuit model.

The low resonant frequency derived based on the EMT is still a little bit counterintuitive considering the fact that the "preconceived" phase accumulations across the double effective plates ($2k_0 h$) and the sandwiched air space ($k_0 d$) are very small. Remember that in a macroscopic sense, there are two well-established physics mechanisms behind the full transmission through a uniform plate: one is the perfect matching of the impedance with surroundings, and the other is the thickness dependent FP resonance. The EOT involved here is closely connected to the latter one. Let's consider an infinite 1D array of the above artificial dielectric plates (with permittivity $e_x = e_0 / h$ and permeability $m_z = h m_0$), arranged periodically (at a pitch $h + d$) along $y$ direction. At the low frequency the system can be further homogenized and corresponds to a uniform medium with effective permittivity $\frac{d e_0 + h e_x}{d + h}$ and permeability $\frac{d m_0 + h m_z}{d + h}$. Again, if the slit is very narrow ($h = 1$), the effective permittivity and permeability can be approximated as $\frac{h e_0}{h(d+h)}$ and $\frac{d m_0}{d+h}$, respectively, resulting in a high relative refraction index $\sqrt{\frac{hd}{h(d+h)^2}}$. As a consequence, the fundamental FP resonant frequency for the current DMG system can be roughly estimated by $\frac{c_0}{4}\sqrt{\frac{h}{hd}}$, which is close to the prediction in Eq. (2). This FP-analogy also clearly elucidates the actual phase change of $p$ exhibited in Fig. 2(c).

As revealed by both the LC-circuit-based microscopic picture and the EMT-based macroscopic model, the resonant frequency $F_R$ is closely relevant to all of the characteristic lengths involved in the DMG system. Besides the decrease of the slit ratio $h = w/p$, both the increase of the SMG thickness $h$ and the separation $d$ enable a drastic reduction of $F_R$. As an example, here we validate the dependence of $F_R$ on $d$, which is easily-controlled for an already fabricated DMG system. As shown in Fig. 4(a), the experimental result (open circle) agrees pretty well with the full-wave calculation (solid line) and the EMT-based analytical model (dash line). It is worth pointing out that although there have been many studies on the transmission response through DMG structures [13-16], which span a wide range of geometric parameters in different frequency regimes (from microwaves to optic waves), no EOT is observed at deep subwavelength (i.e., the resonant wavelength can not resolve the fine features of the structure). Moreover, in previous studies the peak positions are sensitive to the lateral shift, remarkably different from the present case, as shown in Fig. 2.

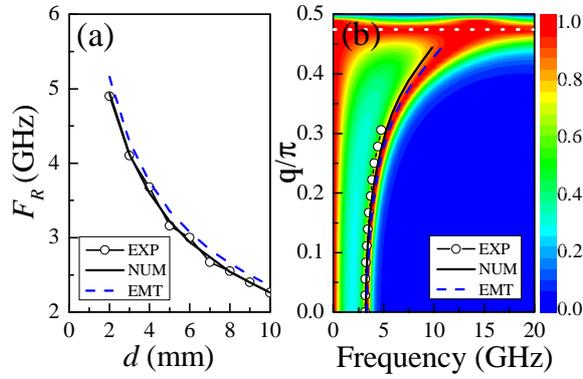

FIG. 4 (color online). (a) The resonant frequency $F_R$ plotted as a function of the separation $d$. (b) The EMT prediction of the angular dependent transmission spectra for a given $d = 5.0$ mm, where the horizontal dot line corresponds to the angular sensitive transparency induced by perfect impedance matching. In (a) and (b) the open circles and solid lines denote respectively the peak positions extracted from the experimental and numerical transmission spectra, while the dash lines come from the EMT-based Eq. (1).

Equation (1) can be straightforwardly extended to the oblique incidence with simply replacing the physics quantities $Z_r$ by $Z_r/\cos q$ and $k_0 d$ by $k_0 d \cos q$, in which the extreme anisotropy ($e_y/e_x \to \infty$) of the artificial dielectric is taken into account. Within a broad range of incident angle $q$, the resonant frequency can be estimated by $F_R \cong \dfrac{c_0}{2p \cos q}\sqrt{\dfrac{2h}{hd}}$, which is still in the deep subwavelength regime as long as the slit ratio $h$ is small enough. This angular robustness is clearly manifested in Fig. 4(b) by the angular dependent transmission spectra evaluated for the EMT-based bilayer system. For comparison, in Fig. 4(b) we also provide the peak positions extracted from the experimental and numerical transmission spectra, which coincide well with the EMT prediction. Note that in Fig. 4(b) the angular spectra also exhibit broadband full transmission near grazing angle (see the horizontal dash line). This attractive complete EM transparency stems from the perfect impedance matching between the metamaterial and surroundings when $\cos q = h$, which has been reported recently in SMG systems and interpreted by an analogy of Brewster-effect [17,18].

In conclusion, we have introduced a new type of EOT effect and elucidated it clearly in both microscopic and macroscopic ways. The theory can be readily extended to the terahertz regime, where the Ohmic loss and plasmonic effects in metals can be neglected. This compelling EOT effect also remains if the space in between the DMG is inserted with solid dielectrics (i.e., silicon oxide) other than the surroundings. Note that the sandwiched structure is quite different from that proposed by Zhou *et al.* [19,20], which involves metamaterial with *negative* permittivity. The compact DMG structure considered here is promising for practical applications because of the merits of angular robustness and excellent immunity to structural imperfections.

**Acknowledgement**

This work was supported by the National Natural Science Foundation of China (Grant Nos. 11174225, 11004155, 11104198); Open Foundation from State Key Laboratory of Applied Optics of China; and the Priority Academic Program Development (PAPD) of Jiangsu Higher Education Institutions.